\documentclass[english,notitlepage,PRL,twocolumn]{revtex4-2}
\usepackage[T1]{fontenc}
\usepackage[latin9]{inputenc}
\usepackage{babel}
\usepackage{amsmath}
\usepackage{graphicx}
\newcommand{\ii}{\mathrm{i}}
\newcommand{\ee}{\mathrm{e}}
\newcommand{\tr}{\mathrm{Tr}}
\newcommand{\sgn}{\mathrm{sgn}}
\newcommand{\secl}[1]{\noindent\textit{#1}}

\begin{document}

\title{Many-body interferometry of one-dimensional integrable systems}

\author{Maksims Arzamasovs}
\thanks{These authors contributed equally to this work.}
\email{marzamasov@xjtu.edu.cn}

\author{Min Liu}
\thanks{These authors contributed equally to this work.}

\author{Yue Zhang}
\author{Rui Tian}
\author{Zehou Li}
\author{Shuai Li}
\author{Bo Liu}
\email{liubophy@gmail.com}
\affiliation{Ministry of Education Key Laboratory for Nonequilibrium Synthesis and Modulation of Condensed Matter,}
\affiliation{Shaanxi Province Key Laboratory of Quantum Information and Quantum Optoelectronic Devices,}
\affiliation{School of Physics, Xi'an Jiaotong University, Xi'an 710049, China}
%\begin{abstract}
%    We propose using many-body Ramsey interferometry to measure non-equilibrium correlation functions of one-dimensional (1D) integrable systems. An archetypal strongly-correlated integrable model, the Ising model is conjectured to equilibrate into non-thermal generalized Gibbs ensemble (GGE) steady states. We outline the experimental protocol for measuring retarded Green's function of the spins in GGE, and use exact numerical calculations to justify the assumption of convergence towards the GGE as the system equilibrates. We show that retarded Green's functions, as opposed to ordinary spin-spin correlators considered previously, can convincingly distinguish between the GGE and thermal post-quench steady-states. We also show that the response functions measurable by Ramsey interferometry can be used to distinguish between different post-quench phases of the model. Our proposal can be realistically realized with modern ultracold atom techniques, and opens up the possibility to study dynamics in exotic non-thermal steady states.
%\end{abstract}

\begin{abstract}
    We propose using many-body Ramsey interferometry to measure non-equilibrium correlation functions of one-dimensional (1D) integrable systems. The 1D transverse-field Ising model, which is conjectured to equilibrate into non-thermal generalized Gibbs ensemble (GGE) steady states, is studied. It is shown that retarded Green's functions, as opposed to ordinary spin-spin correlators considered previously, can convincingly distinguish between the GGE and thermal post-quench steady-states, justifying the assumption of convergence
towards the GGE as the system equilibrates. We also propose the experimental protocol for measuring the response functions with Ramsey interferometry, which can be used to distinguish between different post-quench phases of the model. Our proposal can be realized with current ultracold atom techniques, and opens up the possibility to study dynamics in non-thermal ensembles.
\end{abstract}
\maketitle
As experimental techniques in the field of ultracold atoms are reaching their maturity, the issue of experimental access to the dynamical properties of quantum many-body systems is becoming important and timely. One class of dynamical measurements consists of initializing non-equilibrium time evolution, such as by executing a quantum quench, and then, destructively or non-destructively, tracking relevant observables in time. In fact, a lot of recent works on quantum dynamics have focused on this paradigm, and current state-of-the-art techniques can ensure that this kind of dynamics can be measured very precisely, for example see \cite{Mistakidis_PhysRep_2023} and references therein.

A somewhat different paradigm of dynamical measurements concerns dynamical correlation functions, which are correlated products of time-dependent quantities, but are themselves measured in time-independent steady states, such as the ground state, or the thermal equilibrium. Of particular importance are dynamical response functions, which are the fundamental dynamical quantities that contain information about the spectrum of excitations, and define the dynamical response to external perturbations in the linear response regime. Here even the recent experimental advances can be offset by the conceptual difficulty of non-destructively measuring two or more quantities at different times \cite{Geier_PRXQ_2022}. Measuring dynamical correlation functions requires different approaches from tracking a single time-dependent observable. One technique that has been suggested for effective spin-1/2 systems (which can be realized with two hyperfine states of ultracold alkali atoms) is the many-body Ramsey interferometry. The measurement results in time-resolved retarded Green's functions, which are Fourier-transformed dynamical response functions \cite{Knap_PRL_2013}. This development is interesting because spin-1/2 chains have been a popular choice for research on thermalization in integrable and nearly-integrable models \cite{Vidmar_JSM_2016}, and curiously, there is nothing about the Ramsey technique that would prevent it from being applied in another kind of a diagonal statistical ensemble, beyond the thermal (Gibbs) ensemble, for example the generalized Gibbs ensemble (GGE) which has been proposed and extensively tested for integrable models based on spin-1/2 chains.

This provides a curious premise for a novel kind of dynamical experiment. In this Letter, we propose to measure retarded Green's functions of the spin components,
\begin{equation}
    G^{ab}_{ij}(t_2,t_1)= -i\theta(t_2-t_1)\left\langle S_i^a(t_2) S_j^b(t_1) - S_j^b(t_1) S_i^a(t_2) \right\rangle,
\end{equation}
$a,b\in\{x,y,z\}$, of the finite, integrable Ising chain in a GGE. This would bring three kinds of dynamical concepts together: first, non-equilibrium evolution of the integrable model would be initialized from the grounds state by quenching system parameters. Second, the non-equilibrium system would settle into a non-thermal steady state, described by the non-thermal GGE. Finally, in such a non-thermal but time-independent steady state, time-resolved Green's functions would be measured by Ramsey interferometry. In the following, we introduce the proposed experimental procedure, followed by theoretical discussion based on the exact solution of the time-dependent problem, as well as discussion of experimental feasibility.

\secl{Model.} The first step is initializing the transverse-field Ising model
\begin{equation}
	H=-J\sum_{j=1}^{L-1}S_{j}^{x}S_{j+1}^{x}-h\sum_{j=1}^{L}S_{j}^{y},\label{eq:H_TFI_1}
\end{equation}
in a non-thermal steady state. Being integrable, the Hamiltonian Eq. (\ref{eq:H_TFI_1}) has thermodynamically large number of conserved quantities (for finite systems, "thermodynamically large" means $L$), which we denote by $N_j$,
\begin{eqnarray}
    [H,N_j]=0,\,\,\,\,1\leq j\leq L,
\end{eqnarray}
see theoretical discussion below as well as in \cite{supplemental}, which leads to the breakdown of the eigenstate thermalization hypothesis and makes time evolution non-ergodic. As a result, canonical (Gibbs) ensemble is a possible, but not the most natural equilibrium state for Eq. (\ref{eq:H_TFI_1}), which has lead to the generalized Gibbs ensemble (GGE) being proposed as an alternative \cite{Vidmar_JSM_2016,Ueda_NRP_2020,Mory_JPB_2018,Rylands_AR_2020}, with the density matrix
\begin{equation}
    \rho_{\mathrm{GGE}} = \frac{1}{Z_{\mathrm{GGE}}} \exp\left( -\sum_j \lambda_j N_j\right),
    \label{eq:rho_GGE}
\end{equation}
which maximizes the entropy while taking all $L$ conservation laws into account. Here $Z_{\mathrm{GGE}}=\tr \left(\exp\left(-\sum_j \lambda_j N_j\right)\right)$ is the GGE partition function. Parameters $\lambda_j$ can be seen as generalized "chemical potentials" -- the Lagrangian multipliers in constrained maximization of the entropy, determined from
\begin{equation}
    \left\langle N_i \right\rangle = \tr\left(N_i\rho_{\mathrm{GGE}}\right).
\end{equation}
To initialize Eq. (\ref{eq:H_TFI_1}) in a GGE, the system can be first prepared in its ground state at $J$, $h_0$, followed by quenching the transverse field strength to $h\neq h_0$ at time $t=0$ and allowing the system to equilibrate. Previous numerical studies of the post-quench evolution of Eq. (2) have mostly confirmed the GGE conjecture, but not without unexpected findings. First, spin-spin correlation functions $C^{ab}(t_2,t_1)=|\left\langle S^a(t_2) S^b(t_1)\right\rangle|$ for different $a,b$ were argued to be consistent with either the GGE, or canonical ensemble hypotheses in the same steady state \cite{Rossini_PRL_2009,Rossini_PRB_2010}. However, it is retarded Green's functions $G^{ab}(t_2,t_1)$, rather than spin-spin correlation functions $C^{ab}(t_2,t_1)$ which are accessible via the Ramsey technique, and we show in this work that there is no ambiguity in this case: numerically we can confirm equilibration of correlation functions into a GGE steady state. Second, the case of open boundary conditions was argued to show dramatic finite-size effects, unlike the case of translationally-invariant periodic boundary conditions \cite{Caneva_JSM_2011}. In this work we focused on the OBC case as the most experimentally relevant, and found good convergence properties for time-resolved correlation functions even for finite systems of moderate sizes.

\secl{Many-body Ramsey interferometry.} 
The next and final experimental step is measuring retarded Green's functions Eq. (1) with the many-body Ramsey interferometry technique \cite{Knap_PRL_2013,Cetina_SCI_2016,Ashida_PRB_2018,Kitagawa_PRL_2010,Knap_PRX_2012,Kuklov_PRA_2004}. Recently, Ramsey interferometry has found many uses beyond its historical applications in nuclear magnetic resonance and frequency stabilization of the atomic clock \cite{Ramsey_PR_1950,Vandersypen_RMP_2005,Hahn_PR_1950,Carr_PR_1954}, for example in quantum non-demolition measurements \cite{Haroche_2013} and quantum information and metrology \cite{Cronin_RMP_2009,Degen_RMP_2017,Pezze_RMP_2018}. The technique consists of successively applying two coherent Rabi pulses to the effective spin-1/2 two-level systems:
\begin{equation}
		R_{j}(\theta ,\phi) = 1_j\cos \frac{\theta }{2} + 2\ii\left(S_{j}^{x}\cos\phi - S _{j}^{y}\sin \phi \right)\sin \frac{\theta }{2}.
\end{equation}
Here $1_j$ is the unit operator in the space of spin states at site $j$,  $\theta =\Omega \tau $ with the Rabi frequency $\Omega $ and the pulse duration $\tau $, and $\phi $ is the phase of the laser field. For our purposes we consider spin rotations with $\theta =\pi /2$.
First, a local pulse is applied at time $t_1>t_{eq}>0$ to site $i$ only, corresponding to rotation $R_i(\phi_1)\equiv R_i(\pi/2,\phi_1)$. $t_{eq}$ is introduced as an effective estimate for the time it takes to equilibrate and decay into the steady state described by the GGE density matrix $\rho_{GGE}$. Following the first pulse, the system is allowed to evolve unitarily for time $t$, when the second, global, pulse is applied to all sites at once at time $t_2 = t_1+t$, $R(\phi_2)\equiv \prod_{l=1}^L R_l(\pi/2,\phi_2)$, following which the spin component $S^z_j$ on site $j$ is measured. This results in
\begin{equation}
\begin{split}
   & \langle S^z_j\rangle = M_{ij}(\phi_1,\phi_2,t) = \\
   & \tr \left( R^\dagger_i(\phi_1)\ee^{\ii H t}R^\dagger(\phi_2)S^z_jR(\phi_2)\ee^{-\ii H t}R_i(\phi_1)\rho_{GGE}\right),
\end{split}\label{eq:M_ij}
\end{equation}
which, after some algebra and choosing $\phi_1=0$, $\phi_2=\pi/2$, becomes
\begin{eqnarray}
		\label{eq:M}
%		M_{ij}(\phi _{1},\phi _{2},t) &=&\frac{1}{4}\{\sin (\phi _{1}+\phi
%		_{2})(G_{ij}^{xx}-G_{ij}^{yy}) \nonumber \\
%		&&-\sin (\phi _{1}-\phi _{2})(G_{ij}^{xx}+G_{ij}^{yy})  \nonumber \\
%		&&+\cos (\phi _{1}+\phi _{2})(G_{ij}^{xy}+G_{ij}^{yx})  \nonumber \\
%		&&+\cos (\phi _{1}-\phi _{2})(G_{ij}^{xy}-G_{ij}^{yx})\}  \nonumber \\
 %       &+&\text{terms with odd numbers of }\sigma^{x,y}.
 M_{ij}(0,\pi/2,t) = G^{xx}_{ij,\mathrm{GGE}}(t),
\end{eqnarray}
the GGE Green's function at relative time $t$ (see definition Eq. (12)), where the averages over the terms containing odd numbers of spin operators were ignored due to symmetry considerations, see \cite{supplemental} for discussion of this and other fine points. Thus, for general $t_1,t>0$, $t_1\gg t_{eq}$, generalized Gibbs ensemble Green's function $G_{ij,\mathrm{GGE}}^{xx}(t)$ is obtained as the outcome of the Ramsey protocol, as long as equilibration into the GGE steady state $\rho_{\mathrm{GGE}}$ is assumed, which is can be convincingly proven numerically.

\secl{Exact post-quench evolution.} Owing to the integrability of Eq. (\ref{eq:H_TFI_1}), it is possible to numerically solve systems with hundreds or even thousands of spins by mapping Eq. (\ref{eq:H_TFI_1}) onto a non-interacting fermionic Hamiltonian via the Jordan-Wigner transformations \cite{supplemental, Young_PRB_1997}. Considering an open-boundary (OBC) system containing $L=600$ spins, we implemented three ways of computing the post-quench retarded Green's functions to see which stationary ensemble the true Green's function $G^{xx}_{ij,\mathrm{U}}(t_2,t_1)$ relaxes to in the limit of $t_1\gg t_{eq}$. Using the Heisenberg picture formalism, exact post-quench Green's functions are
\begin{equation}
    G^{xx}_{ij,\mathrm{U}}(t_2,t_1)=-\ii\left\langle\psi_{init}\right| \left[ S^x_i(t_1), S^x_j(t_1+t)\right] \left|\psi_{init}\right\rangle, \label{eq:ret_GF_unitary}
\end{equation}
where the subscript "U" stands for unitary evolution, $\left|\psi_{init}\right\rangle$ is the pre-quench ground state, and
\begin{equation}
    S^x_j(t_1) = e^{iH t_1} S^x_j e^{-iH t_1}
\end{equation}
\begin{equation}
    S^x_j(t_2) = S^x_j(t_1 + t) = e^{iH(t+t_1)} S^x_j e^{-iH(t+t_1)}
\end{equation}
are the time-dependent Heisenberg operators, evolved with the post-quench Hamiltonian. These are the exact post-quench Green's functions, without any reference to the existence or the possible nature of the post-quench steady-state. On the other hand, steady-state Green's functions in the assumed GGE are
\begin{equation}
    G^{xx}_{ij,\mathrm{GGE}}(t)=-\ii\tr\left(\left[ S^x_i(t_1), S^x_j(t_1+t)\right]\rho_{\mathrm{GGE}}\right) ,\label{eq:ret_GF}
\end{equation}
Since $\rho_{\mathrm{GGE}}$ is time-independent, $G^{xx}_{ij,\mathrm{GGE}}(t_2,t_1)\equiv G^{xx}_{ij,\mathrm{GGE}}(t)$ does not actually depend on $t_1$ and $t_2$ separately, only on the time difference $t=t_2-t_1$. Values of $\lambda_j$ are inferred from $J$, $h_0$, and $h$, using Eq. (5) with $\left\langle N_j \right\rangle=\left\langle\psi_{init}\right| N_j \left|\psi_{init}\right\rangle$, see \cite{Damski_JPA_2014,Vidmar_JSM_2016,supplemental}. Finally, in view of the previously reported ambiguity \cite{Rossini_PRB_2010}, we also implemented the canonical (Gibbs) ensemble for comparison,
\begin{equation}
    G^{xx}_{ij,\mathrm{G}}(t)=-\ii\tr\left(\left[ S^x_i(t_1), S^x_j(t_1+t)\right]\rho_{\mathrm{G}}\right) ,\label{eq:ret_GF_thermal}
\end{equation}
where "G" stands for Gibbs, and
\begin{equation}
    \rho_\mathrm{G}  = \frac{1}{Z_\mathrm{G}}\exp\left(-\frac{H}{T_{eff}}\right), \label{rho_GE}
\end{equation}
$Z_\mathrm{G} = \tr\left(\exp\left(-H/T_{eff}\right)\right)$, is the canonical (Gibbs, thermal) partition function. $T_{eff}$ is the effective temperature, fixed by the conservation of energy
\begin{equation}
    \left\langle\psi_{init}\right| H \left|\psi_{init}\right\rangle = \tr\left(H\rho_{\mathrm{G}}\right).
\end{equation}
Details of the calculation are given in \cite{supplemental}, and the findings are summarized below.

We were able to unambiguously show that post-quench Green's functions $G_\mathrm{U}$ converge towards the GGE predictions $G_{\mathrm{GGE}}$. In Fig. 1 we show convergence of the same-time, nearest-neighbor correlation function as a function of time elapsed after the quench, $t_1$. This is shown as a proof of principle, and also as a comment on reported differences in convergence for open and periodic transverse Ising systems \cite{Caneva_JSM_2011}. In our calculations, we find nothing particular about the OBC case. On the contrary, we find excellent convergence towards the GGE predictions for systems of even smaller sizes, such as $L=200$, as well as excellent finite-size scaling. For the system at hand with L=600, convergence is achieved for $t_1 J>10$, which sets the scale for $t_{eq}$. In Fig. 2 we establish convergence of the exact unitary evolution result Eq. (9) towards the GGE prediction Eq. (12) at long $t_1$. Namely, we study time-dependent response functions, Eqs. (12) and (13), and compare them to the exact unitary evolution result Eq. (9) at late $t_1$, as the system relaxes to a steady state. We indeed observe excellent convergence towards the GGE result. In other words, it is justified to say that, at long $t_1$, the system approaches the GGE. On the other hand, there are obvious discrepancies between the curves corresponding to the exact unitary evolution and the canonical ensemble.
\begin{figure}[!ht]
	\includegraphics[width=0.5\textwidth]{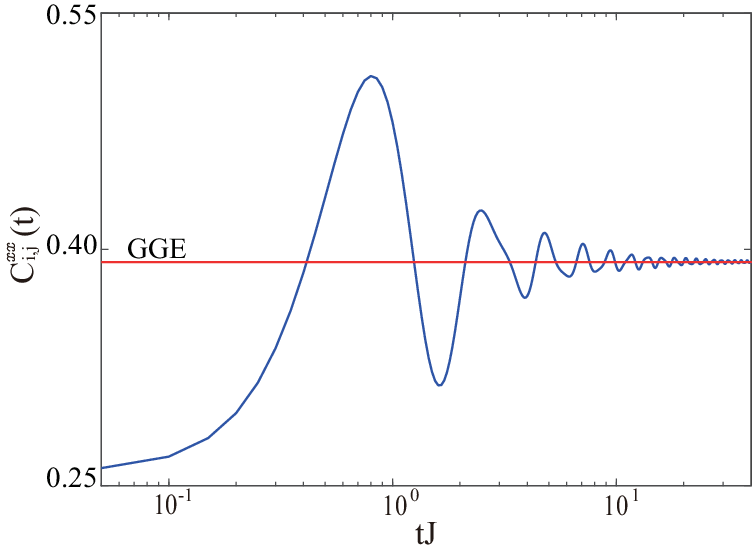}\label{fig:1}
	\caption{Post-quench same-time correlation function $C^{xx}_{ij,\mathrm{U}}(t_1,t_1)=|\langle S^x_i(t_1) S^x_j(t_1) \rangle|$ with L=600,i=j-1=300, $h_0=2J$, $h=0.3J$ under the OBC. Red horizontal line denotes the corresponding GGE result, $C^{xx}_{ij,\mathrm{GGE}}(0)$.}
\end{figure}
\begin{figure}[!ht]
	\includegraphics[width=0.5\textwidth]{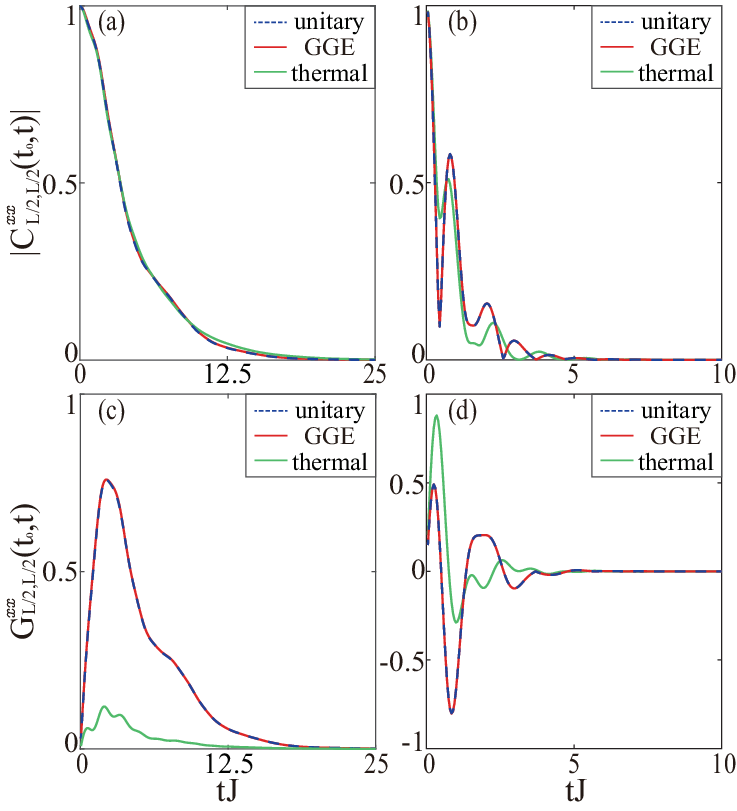}\label{fig:2}
	\caption{(a)(b) Correlation function and (c)(d) Green's function with L=600 under the OBC. The solid red line represents unitary time evolution, the dashed blue line represents generalized Gibbs ensemble, and the solid green line represents thermal ensemble. $t_{1}=20J$ in the case of unitary evolution, and GGE and canonical Green's functions are independent of $t_1$. Quench parameters are: (a)(c) $h_{0}=2J$, $h=0.3J$ (paramagnetic to ferromagnetic) and (b)(d) $h_{0}=0.3J$, $h=2J$(ferromagnetic to paramagnetic).}
\end{figure}
It is interesting to note that non-retarded spin-spin correlation functions of the Ising model
\begin{equation}
    C^{ab}_{ij,\mathrm{U}}(t_2,t_1)= |\left\langle\psi_{init}\right| S^a_i(t_1) S^b_j(t_2) \left|\psi_{init}\right\rangle| ,\label{eq:ret_CF_unitary}
\end{equation}
\begin{equation}
    C^{ab}_{ij,\mathrm{GGE}}(t)=| \tr\left( S^a_i(t_1) S^b_j(t_1+t)\rho_{\mathrm{GGE}}\right)| ,\label{eq:ret_CF_GGE}
\end{equation}
\begin{equation}
    C^{ab}_{ij,\mathrm{G}}(t)=|\tr\left( S^a_i(t_1) S^b_j(t_1+t)\rho_{\mathrm{G}}\right)| ,\label{eq:ret_CF_GE}
\end{equation}
had been studied previously \cite{Rossini_PRL_2009,Rossini_PRB_2010}, and it had been argued that the post-quench steady state achieved by Eq. (\ref{eq:H_TFI_1}) was, quite peculiarly, showing signatures of both canonical and GGE ensembles, which was manifested by accessing different correlation functions: $C^{xx}_{ij}$ was claimed to show thermal behavior governed by $\rho_\mathrm{\mathrm{G}}$, whereas late-time behavior of $C^{yy}_{ij}$ was claimed to be governed by $\rho_{\mathrm{GGE}}$. We studied convergence of Eqs. (16)-(18) at large $t_1$ in the upper panels of Fig. 2 and found that, for the spin-spin correlators and Green's functions we considered, it was not necessary to invoke the Gibbs ensemble at all: it turns out that, even though the correlation function $C^{xx}_{ij,\mathrm{U}}$ is reasonably approximated by the canonical ensemble average at long $t_1$, it is described by the GGE the best, still. Most interestingly, when looking at convergence of the retarded Green's functions, plotted in the lower panels of Fig. 2, the canonical ensemble approximation is visibly different in many cases, whereas the GGE average matches the unitary evolution result at late $t_1$ almost perfectly. Thus the canonical ensemble hypothesis seems to be of limited use, and we end up confirming the original, simpler, conjecture that the post-quench steady state of the transverse Ising model Eq. (\ref{eq:H_TFI_1}) is described by the GGE Eq. (\ref{eq:rho_GGE}), as manifested by both the correlation and retarded Green's functions, 

%In both cases, correlation functions $C^{xx}_{ij}$ have been shown \cite{Sachdev_PRL_1997,Rieger_PRB_2011} to decay exponentially as
%\begin{equation}
%    C^{xx}_{i,i+r}(t)\sim\exp\left(-\int_{\pi}^{\pi}\frac{dk}{\pi}\langle N_{k}\rangle\left|r-v_{k}t\right|\right),
%\end{equation}
%where $k$ are quasiparticle momenta (which are good quantum numbers in the large $L$ limit), $\langle N_k \rangle$, $\epsilon_k$, and $v_k=d\epsilon_k/dk$ are the quasiparticle occupation numbers in the generalized Gibbs ensemble, dispersion, and velocity, respectively. The paramagnetic phase also has the periodic modulation superimposed on top of the decay \cite{supplemental}.

\secl{Physical significance of GGE Green's functions.} Experimentally realizing the GGE and measuring the GGE Green's functions is interesting in and of itself, and can enhance our understanding of quantum dynamics beyond the thermal equilibrium, such as by extending the linear response theory to the generalized Gibbs ensembles \cite{Mahan}. Here we present another observation regarding quenches between different phases of Eq. (\ref{eq:H_TFI_1}), offering another possible use. 

The transverse Ising model exhibits a quantum phase transition at $|h|=|J|$, separating the long-range ordered (ferromagnetic for $J>0$, antiferromagnetic for $J<0$, low $|h|$) and the polarized (paramagnetic, high $|h|$) phases \cite{CherngLevitov}. Therefore, in addition to quantitative studies of the convergence properties of the retarded Green's function, we also looked at more qualitative behaviors in their dependence on the parameters of the quench, specifically when $h$ and $h_0$ belong to different phases. In Fig. 3 we show two results for the GGE Green's function, initialized by quenches across the phase transition point. The behavior of post-quench correlation functions, $C^{xx}_{ij}(t_2,t_1)$, has previously been analyzed based on the semi-classical picture, both for periodic and open boundary conditions \cite{Sachdev_PRL_1997,Rieger_PRB_2011}. Quenches into both (anti) ferromagnetic and paramagnetic phases create free particle-like excitations above the corresponding ground states: in the (anti) ferromagnetic case, the quasiparticles are the domain walls in the $x$ direction, and in the paramagnetic case, the quasiparticles are single flipped spins in the $y$ direction. These quasiparticles are nothing but the fermionic quasiparticles diagonalizing the final Hamiltonian, with their occupation numbers determined by the nature of the steady state (i.e. thermal or GGE), and the initial and post-quench values of $h$. Using classical probabilistic arguments for the trajectories of the semi-classical particles, expressions for spin-spin correlation functions have been derived, resulting in an exponentially decaying shape for quenches into the ferromagnetic phase, and exponentially decaying shape superimposed with oscillations for quenches into the paramagnetic phase \cite{Sachdev_PRL_1997,Rieger_PRB_2011}. 

Interestingly, our results for the retarded Green's function $G_{ij,\mathrm{GGE}}$, which can be thought of as the imaginary part of the correlation function $C_{ij,\mathrm{GGE}}$, are in qualitative agreement with these predictions as well. This can be explained qualitatively by arguing that the steady-state $\left\langle S^x(0)S^x(t)\right\rangle$ equals $\left\langle S^x(t')S^x(0)\right\rangle$ evaluated at the negative time $t'=-t$, hence the difference of $\left\langle S^x(0)S^x(t)\right\rangle$ and $\left\langle S^x(t)S^x(0)\right\rangle$, which is proportional to $G^{xx}_{ij,\mathrm{GGE}}(t)$, shows the same qualitative long-time behavior as each of the correlation functions separately -- either decay or decay superimposed with oscillations. Thus, measuring post-quench response functions can also qualitatively probe the phase of the post-quench transverse field Ising model by looking at the presence or absence of oscillations. This may serve as another example of using post-quench dynamics to classify equilibrium phases, similar to identifying topological phases by measuring the Hopf index following a quantum quench \cite{Wang_PRL_2017,Sun_PRL_2018}.
\begin{figure}[]
	\includegraphics[width=0.5\textwidth]{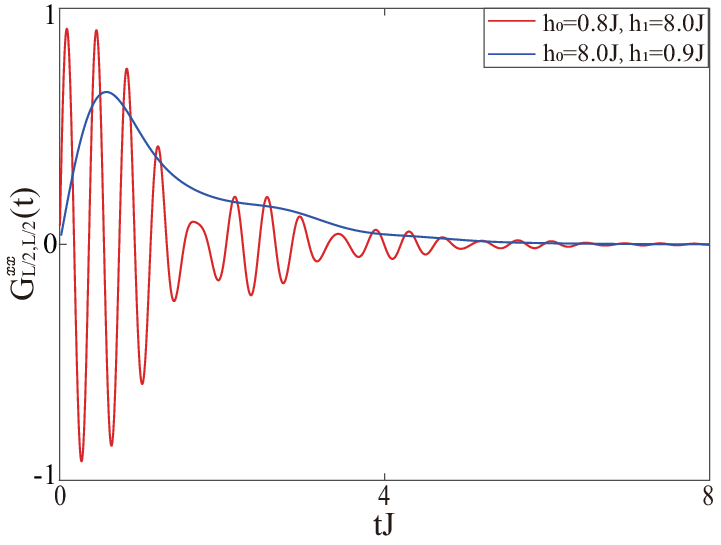}\label{fig:3}
	\caption{GGE Green's functions for L=600 under the OBC for two different quenches: from the paramagnetic to ferromagnetic phase (blue line), and from the ferromagnetic to paramagnetic phase (red line). }
\end{figure}

\secl{Experimental feasibility.} Our proposal should be realizable with current ultracold techniques. Indeed, several experimental proposals for realizing spin models with ultracold atoms have been recently put forward \cite{Liu_PRL_2018,Schauss_QST_2018,Liao_PRA_2021,Liao_PRA_2022}, with antiferromagnetic nearest-neighbor Ising model actually realized in experiments \cite{Simon_Nature_2011}. Other options include simulating spin chains with Rydberg atoms \cite{Zeiher_NatPhys_2016} or with strongly-interacting bosons \cite{Barfknecht_PRR_2021,Mistakidis_PhysRep_2023}.

Ramsey-interferometric measurements of time-resolved correlation functions offer several advantages over the alternatives: a single projective measurement following the two Rabi pulses allows to read off space- and time-resolved Green's function exactly, whereas an alternative of performing several subsequent weak measurements at different points in time would potentially destroy the coherence \cite{Geier_PRXQ_2022}. And even though spectroscopic techniques, such as the Bragg scattering, can access frequency- and momentum-dependent response functions, reliable mapping into time and space domains by the means of the Fourier transform is often undermined by experimental limitations on the available bandwidth \cite{Knap_PRL_2013}, making the ultracold atom implementation of Ramsey interferometry the most direct way of experimentally initializing and studying time-resolved correlators, including in the generalized Gibbs ensemble. For some other proposals for measuring various time-resolved correlation functions in and out of equilibrium based on Ramsey interferometry and state-of-the-art ultracold techniques see \cite{Koller_PRL_2014,Uhrich_PRA_2017,Uhrich_QST_2019,Kastner_PRA_2020, Alhambra_PRL_2020, Castrignano_PRA_2020, Baez_PNAS_2020}.

\secl{Discussion and conclusions.} To summarize, we have put forward a proposal for measuring retarded spin-spin correlation functions of the nearest-neighbor transverse-field Ising model in a non-thermal GGE steady state, which is the most natural equilibrium state for a class of integrable models. To that end, we proposed extending the many-body Ramsey protocol of \cite{Knap_PRL_2013} to the case of the generalized Gibbs ensemble, beyond its current applications in thermal equilibrium. By numerically tracking the evolution of time-resolved correlation functions following a quantum quench, we have shown that, at late times, the OBC systems unambiguously reach GGE steady states (and do not reach the canonical (thermal) ensemble steady states, as opposed to an exotic superposition of the canonical and generalized Gibbs ensembles which was previously conjectured). Finally, we have observed that the post-quench retarded Green's functions have qualitative features which can be useful in identifying the post-quench phase of the Ising model by the presence or absence of oscillations.

\secl{Acknowledgments} This work is supported by the National Key R$\&$D Program of China (2021YFA1401700), NSFC (Grants No. 12074305, 12474267), the Fundamental Research Funds for the Central Universities (Grant No. xtr052023002), the Shaanxi Fundamental Science Research Project for Mathematics and Physics (Grant No. 23JSZ003) and the Xiaomi Young Scholar Program. We also thank the HPC platform of Xi'an Jiaotong University, where our numerical calculations was performed.
\bibliography{GGE_bib}

\onecolumngrid
\appendix

\section{Exactly solving the nearest-neighbor Ising Hamiltonian with open boundary conditions} 
Hamiltonian Eq. (2) can be recast as the free fermionic quasiparticle Hamiltonian by applying the Jordan-Wigner transformations,
\begin{eqnarray}
		S_{j}^{x} - \ii S_{j}^{z}  &=&c_{j}^{\dagger }\exp \left( -\ii\pi
		\sum_{l=1}^{j-1}c_{l}^{\dagger }c_{l}\right) ,\, \notag  \\
		S_{j}^{x} + \ii S_{j}^{z} &=&c_{j}\exp \left( \ii\pi \sum_{l=1}^{j-1}c_{l}^{\dagger
		}c_{l}\right) ,\,\notag \\
		S_{j}^{y} &=&c_{j}^{\dagger }c_{j}-\frac{1}{2},\label{eq:JW_supplemental}
\end{eqnarray}
which turns Eq. (\ref{eq:H_TFI_1}) of the main text into Fermionic Hamiltonian with pairing terms
\begin{equation}
H =-\frac{J}{4}\sum_{j=1}^{L-1}\left( c_{j}^{\dagger
}c_{j+1}+c_{j+1}^{\dagger }c_{j}\right)-\frac{J}{4}\sum_{j=1}^{L-1}\left( c_{j}^{\dagger }c_{j+1}^{\dagger
}+c_{j+1}c_{j}\right) -h\sum_{j=1}^{L}c_{j}^{\dagger }c_{j}+\frac{hL}{2},\label{eq:transverse_Ising}
\end{equation}
where $J>0$ is assumed, i.e. the ordered phase $|h|<|J|$ is ferromagnetic rather than anti-ferromagnetic. The constant term $hL/2$ is customarily omitted and the remainder of Eq. (A2) can be arranged as
\begin{equation}
    H=\left(\begin{array}{cccccc}
c_{1}^{\dagger} & ... & c_{L}^{\dagger} & c_{1} & ... & c_{L}\end{array}\right)H^{sp}\left(\begin{array}{c}
c_{1}\\
...\\
c_{L}\\
c_{1}^{\dagger}\\
...\\
c_{L}^{\dagger}
\end{array}\right),
\end{equation}
where $H^{sp}$ is a real, symmetric $2L\times 2L$ single-particle Hamiltonian matrix, which can be diagonalized by an orthogonal transformation $V$,
\begin{equation}
    H^{sp} = V^T D V,
\end{equation}
where $D$ is diagonal and is made up of $L$ positive-negative pairs $\pm \epsilon_j$, $1 \leq j \leq L$. Explicitly choosing the positive eigenvalues, $\epsilon_j>0$, the Jordan-Wigner transformed Hamiltonian Eq. (A2) can be rearranged as
\begin{equation}
    H = \sum_{j=1}^L \epsilon_j \gamma^\dagger_j \gamma_j - \epsilon_j \gamma_j \gamma^\dagger_j + \frac{hL}{2} = \sum_{j=1}^L \epsilon_j \left(2\gamma^\dagger_j \gamma_j - 1\right) + \frac{hL}{2},
    \label{eq:transverse_Ising_diag}
\end{equation}
with $\gamma^\dagger_j$ and $\gamma_j$ being Fermionic quasiparticle creation and annihilation operators \cite{Vidmar_JSM_2016,Young_PRB_1997}. Where it does not alter the physics, we are going to ignore constant terms and write $H=\sum_{j=1}^L 2\epsilon_j \gamma^\dagger_j \gamma_j$. For open boundary conditions (OBC), index $1\leq j \leq L$ enumerates the quasiparticle energy levels and does not stand for any particular quantum number. For periodic boundary conditions (PBC), the quasiparticles can be enumerated by momentum $-\pi\leq k<\pi$, and analytic expressions for $\epsilon_k$, and $\gamma_k$ in terms of $c_j$, $c^\dagger_j$ are available \cite{Damski_JPA_2014}. However, the periodic case comes with its subtleties \cite{Damski_JPA_2014}. In this work we talk about potential experimental applications, so the focus is primarily on the OBC case, and only in Appendix F we invoke the PBC results since in the limit of sufficiently large $L$, and far from the boundaries, convergence between PBC and OBC results is expected.

\section{Density matrix of the generalized Gibbs ensemble} 
The model Eq. (2) (alternatively, Eq. (A5)) has $L$ conserved quantities $N_j=\gamma_j^\dagger\gamma_j$, $[H,N_j]=0$, which do not change during the time evolution and can be set independently. Thus, ergodicity is broken and an isolated Ising model cannot thermalize. Instead, generalized Gibbs ensemble (GGE) density matrix has been proposed as a steady state that maximizes the entropy while satisfying all $L$ conservation laws,
\begin{equation}
\rho _{\mathrm{GGE}} =\frac{1}{Z_{\mathrm{GGE}}}\exp \left( -\sum_{j}\lambda _{j}N_{j}\right)
= \frac{\prod_{j}\exp \left( -\lambda _{j}\gamma _{j}^{\dagger
}\gamma _{j}\right) }{\prod_{j}\tr[\exp \left( -\lambda _{j}\gamma
_{j}^{\dagger }\gamma _{j}\right) ]},
\end{equation}%
where $\lambda _{j}$ are numerical parameters found by fitting the GGE averages $\langle N_{j}\rangle$ to their actual expectation values. For example, quenching from the initial ground state of the pre-quench Hamiltonian $|\psi _{\mathrm{GS}}\rangle$, conservation laws require
\begin{equation}
\langle N_{j}\rangle =\langle \psi _{\mathrm{GS}}|N_{j}|\psi _{\mathrm{GS}}\rangle =\tr%
(N_{j}\rho _{\mathrm{GGE}}),
\label{eq:GGE_initial_condition}
\end{equation}
which leads to \cite{Vidmar_JSM_2016}
\begin{equation}
    \lambda_j = \ln\left( \frac{1- \langle N_j  \rangle}{\langle N_j  \rangle} \right).
\end{equation}
In our subsequent calculations we apply Wick's theorem to products of operators $c^\dagger_j$ and $c_j$, therefore computing the values of $\lambda_j$ is not necessary, and it suffices to know $\langle N_j\rangle$, which are fixed by the initial conditions Eq. (\ref{eq:GGE_initial_condition}).

\section{Canonical ensembles with effective temperature $T_{eff}$}
%As mentioned above, although $\lambda_i$ or $T$ is not needed in the calculation of Green's function in the ensemble, but only $\langle n_j\rangle$, it is necessary to introduce them.\\
%Observables in the GGE, Eq. (\ref{eq:rho_GGE}) is the crucial density matrix. The corresponding partition function is $Z_{\rm GGE} = {\rm Tr}[e^{-\sum_k \lambda_k \hat I_k}]$. The Lagrange multipliers $\{ \lambda_k \}$ are fixed by the initial conditions $\langle I_k \rangle_0 = \langle \psi_0 |\hat I_k | \psi_0 \rangle =  {\rm Tr} [ \hat \rho_{\rm GGE}  \hat I_k]$ \cite{Vidmar_JSM_2016}. We denote the expectation values of observables in the GGE as $\langle\hat{\cal O}\rangle_{\rm GGE}\equiv\textrm{Tr}[\hat\rho_{\rm GGE}\hat{\cal O}]$. Take  XX model as an example, the partition function of the GGE can be written as 
%\begin{equation}
%Z_{\rm GGE} = {\rm Tr}[  e^{-\sum_q \lambda_q \hat I_q}] = \prod_q (1+e^{-\lambda_q}),
%\end{equation}
%where $q=1,2,\ldots,L$. The expectation values of conserved quantities in the GGE can be written as
%\begin{equation}
%\langle \hat I_q  \rangle_{\rm{GGE}} = Z_{\rm GGE}^{-1} {\rm Tr} [ e^{-\sum_q \lambda_q \hat I_q}  \hat I_q] = %\frac{e^{-\lambda_q}}{1+e^{-\lambda_q}}.
%\end{equation}
%Since the GGE is constructed requiring that those expectation values are the same as in the initial state $\langle \hat I_q  \rangle_0= \langle \psi_0 |\hat I_q | \psi_0 \rangle$, one obtains the following expression for the Lagrange multipliers
%\begin{equation} 
%\lambda_q = \ln\left( \frac{1- \langle \hat I_q  \rangle_0}{\langle \hat I_q  \rangle_0} \right),
%\end{equation}
Even though breakdown of ergodicity prevents an isolated Ising model from reaching thermal equilibrium, for the sake of comparison we do evaluate hypothetical thermal Green's and correlation functions. The effective temperature $T_{eff}$ is determined from the conservation of energy by matching its values in the initial pure post-quench state $|\psi _{GS}\rangle$ to that of the (assumed) eventual Gibbs ensemble,
\begin{equation}
    \langle E \rangle = 2\,\tr\left(\sum_j \epsilon_j \gamma^\dagger_j \gamma_j\rho_\mathrm{G}\right)= 2\langle\psi_{GS}|\sum_j \epsilon_j \gamma^\dagger_j \gamma_j|\psi_{GS}\rangle,
\end{equation}
with the thermal average given by
\begin{equation}
    \langle E \rangle = 2\,\tr\left(\sum_j \epsilon_j \gamma^\dagger_j \gamma_j\rho_\mathrm{G}\right)=\sum_{j=1}^L \frac{2\epsilon_j}{\exp\left(2\epsilon_j/T_{eff}\right)+1},
    \label{eq:av_energy_thermal}
\end{equation}
with $\epsilon_j$ being the post-quench quasiparticle energies. The initial value $\langle E \rangle = 2\langle\psi_{\mathrm{GS}}|\sum_j \epsilon_j \gamma^\dagger_j \gamma_j|\psi_{\mathrm{GS}}\rangle$ is found from the knowledge of pre- and post-quench values of the magnetic field $h_0$ and $h$.
 
\section{Correlation functions of the spins}

Evaluating retarded Green's functions $G^{xx}_{ij}$ requires computing the average of $\left[ S^x_i(t_1+t), S^x_j(t_1)\right] = S^x_i(t_1+t)S^x_j(t_1) - S^x_j(t_1)S^x_i(t_1+t)$. Below we use the technique outlined in \cite{Young_PRB_1997} to demonstrate how $S^x_i(t_1+t)S^x_j(t_1)$ is calculated, and the other term can be treated similarly. Expanding the Jordan-Wigner strings as
\begin{equation}
\prod_l\exp \left( i\pi c_{l}^{\dagger }c_{l}\right) =\prod_l\left( c_{l}^{\dagger
}+c_{l}\right) \left( c_{l}^{\dagger }-c_{l}\right),
\end{equation}
and introducing 
\begin{equation}
    A_{l}=c_{l}^{\dagger }+c_{l},\quad A^2_{l}=1, \notag
\end{equation}
\begin{equation}
    B_{l}=c_{l}^{\dagger }-c_{l},\quad B^2_{l}=-1,
\end{equation}
results in
\begin{equation}
S_{i}^{x}(t_1+t)S_{j}^{x}(t_1)=\frac{1}{4}\prod_{m=1}^{i-1}\left(
A_{m}(t_1+t)B_{m}(t_1+t)\right) A_{i}(t_1+t)\prod_{l=1}^{j-1}\left(
A_{l}(t_1)B_{l}(t_1)\right) A_{j}(t_1),  \label{eq:ensemble}    
\end{equation}
where $A_i(t_1)$, $B_i(t_1)$, etc, are the Heisenberg versions of operators Eq. (D2). For products of Fermionic operators $A_i$ and $B_i$, Wick's theorem applies to averaging performed in 1) canonical, 2) generalized Gibbs ensembles, or 3) in the pre-quench ground state. The result is a Pfaffian of the matrix of all possible pairwise averages, which can be recast as a square root of the determinant of a skew-symmetric matrix, up to the sign ambiguity \cite{Young_PRB_1997}. Namely, if
\begin{equation}
ABC...Z
\end{equation}
is a product of Fermionic operators $A_j$ or $B_j$, then
\begin{equation}
\left\langle ABC...Z\right\rangle^2 =\left\vert 
\begin{array}{ccccc}
0 & \left\langle AB\right\rangle  & \left\langle AC\right\rangle  & \ldots 
& \left\langle AZ\right\rangle  \\ 
-\left\langle AB\right\rangle  & 0 & \left\langle BC\right\rangle  & \ldots 
& \left\langle BZ\right\rangle  \\ 
-\left\langle AC\right\rangle  & -\left\langle BC\right\rangle  & 0 & \ldots 
& \left\langle CZ\right\rangle  \\ 
\vdots  & \vdots  & \vdots  & \ddots  & \vdots  \\ 
-\left\langle AZ\right\rangle  & -\left\langle BZ\right\rangle  & 
-\left\langle CZ\right\rangle  & \ldots  & 0%
\end{array}%
\right\vert .
\end{equation}%

In time-independent, stationary canonical or generalized Gibbs ensembles, $t_1$ can be set to zero, which allows to simplify $\langle A_i(0) A_j(0) \rangle = \langle A_i(t) A_j(t) \rangle = \delta_{ij}$, $\langle B_i(0) B_j(0) \rangle = \langle B_i(t) B_j(t) \rangle = -\delta_{ij}$. Additionally, we encounter the following averages: $\langle A_i(t)A_j(0)\rangle$, $\langle B_i(t)B_j(0)\rangle$, $\langle B_i(t)A_j(0)\rangle$, $\langle A_i(t)B_j(0)\rangle$, which evaluate to

\begin{eqnarray}
\langle A(t)A(0)\rangle  &=&\psi \left( U(t)+V(t)\right) \psi ^{\dagger },\, 
\notag \\
\langle B(t)B(0)\rangle  &=&\phi \left( -U(t)-V(t)\right) \phi ^{\dagger }, 
\notag \\
\langle A(t)B(0)\rangle  &=&\psi \left( -U(t)+V(t)\right) \phi ^{\dagger }, 
\notag \\
\,\langle B(t)A(0)\rangle  &=&\phi \left( U(t)-V(t)\right) \psi ^{\dagger }.
\end{eqnarray}%
where matrices $\psi$ and $\phi$ are determined by the expansion coefficients between operators $c$, $c^\dagger$ and quasiparticle operators of the post-quench Hamiltonian,
\begin{equation}
c_{i}^{\dagger }+c_{i}=\sum\limits_{\mu }\psi _{\mu i}(\gamma _{\mu
}^{\dagger }+\gamma _{\mu }),\quad c_{i}^{\dagger }-c_{i}=\sum\limits_{\mu }\phi
_{\mu i}(\gamma _{\mu }^{\dagger }-\gamma _{\mu }),
\end{equation}
and
\begin{eqnarray}
V(t) &=&\mathrm{diag}(1-\langle N_{j}\rangle \exp (-\mathrm{i}\epsilon
_{j}t)),  \notag \\
U(t) &=&\mathrm{diag}(\langle N_{j}\rangle \exp (-\mathrm{i}\epsilon
_{j}t)),
\end{eqnarray}
where $\langle N_{j}\rangle = \langle \gamma^\dagger_j \gamma_j \rangle$ are the post-quench quasiparticle occupation numbers evaluated in the corresponding ensemble, either canonical or generalized Gibbs.

For averages following the exact unitary evolution of the pre-quench ground state,
\begin{equation}
\langle S_{i}^{x}(t+t_1)S_{j}^{x}(t_1) = \langle \psi _{\mathrm{GS}}|e^{\ii H(t+t_{1})}S_{i}^{x}e^{-\ii %
H(t+t_{1})}e^{\ii H t_{1}}S_{j}^{x}e^{-\ii Ht_{1}}|\psi
_{\mathrm{GS}}\rangle,
\label{corr-fun-unt}
\end{equation}
with $|\psi_{\mathrm{GS}}\rangle$ being the ground state of the pre-quench Hamiltonian $H_0$, the pairwise averages are 
\begin{eqnarray}
\langle A(t_1)A(t_2)\rangle  &=&\frac{1}{2}\psi \ee^{2\ii 
D t_1}V^{\dagger }\left( 1-\sgn(H^{sp}_0)\right) V\ee^{-2\ii 
D t_2}\psi^\dagger,  \notag \\
\langle A(t_1)B(t_2)\rangle &=&-\frac{1}{2}\psi \ee^{2\ii D (t_1)}V^{\dagger }\left( 1-\sgn(H^{sp}_0)\right) V\ee^{-2\ii D t_2} \phi^\dagger, \notag \\
\langle B(t_1)A(t_2)\rangle  &=&\frac{1}{2}\phi \ee^{2\ii 
D (t_1)}V^{\dagger }\left( 1-\sgn(H^{sp}_0)\right) V\ee^{-2\ii D t_2}\psi^\dagger,  \notag \\
\langle B(t_1)B(t_2)\rangle  &=&-\frac{1}{2}\phi \ee^{2\ii D (t_1)}V^{\dagger }\left( 1-\sgn(H^{sp}_0)\right) V\ee^{-2\ii D t_2}\phi^\dagger.
\end{eqnarray}
where $t_1,t_2$ are $t+t_1$ or $t_1$, $H^{sp}_0$ is the single-particle Hamiltonian corresponding to the pre-quench Hamiltonian $H_0$,
\begin{equation}
    H_0 = \left(\begin{array}{cccccc}
c_{1}^{\dagger} & c_{2}^{\dagger} & \ldots & c_{1} & c_{2} & \ldots\end{array}\right)H^{sp}_{0}\left(\begin{array}{c}
c_{1}\\
c_{2}\\
\ldots\\
c_{1}^{\dagger}\\
c_{2}^{\dagger}\\
\ldots
\end{array}\right),
\end{equation}
and similarly for post-quench Hamiltonian $H$, and $V_0$, $V$ are the matrices diagonalizing $H_0$, $H$,
\begin{equation}
    V^\dagger_0 H^{sp}_0 V_0 = D_0,\quad V^\dagger H^{sp} V = D.
\end{equation}
Finally, $\sgn(...)$ of a matrix is the shorthand for
\begin{equation}
    \sgn(H^{sp}_0) = V_0\,\sgn(D_0)\,V_0^\dagger.
\end{equation}

\section{Retarded Green's functions from Ramsey interferometry}
Full version of Eq. (\ref{eq:M}) of the main text is 
\begin{eqnarray}
M_{ij}(\phi _{1},\phi _{2},t)&=&\frac{1}{2}\{\sin (\phi _{1}+\phi _{2})(G_{ij}^{xx}-G_{ij}^{yy}) - \sin (\phi _{1}-\phi _{2})(G_{ij}^{xx}+G_{ij}^{yy}) + \cos (\phi _{1}+\phi _{2})(G_{ij}^{xy}+G_{ij}^{yx})  \notag \\
&&+\cos (\phi _{1}-\phi _{2})(G_{ij}^{xy}-G_{ij}^{yx})\}+\frac{1}{2}\langle 
\hat{C}(t) + 4(S_{i}^{x}\cos \phi _{1}-S _{i}^{y}\sin \phi _{1})\hat{C}%
(t)(S_{i}^{x}\cos \phi _{1}-S _{i}^{y}\sin \phi _{1})\rangle  
\end{eqnarray}
where $\hat{C}(t)=-[S_{j}^{x}(t)\sin \phi _{2}+S_{j}^{y}(t)\cos \phi
_{2}]$. Further choosing
\begin{equation}
    \phi_1=0,\quad\phi_2=\frac{\pi}{2},
\end{equation}
only the following odd product averages survive:
\begin{equation}
    -\langle\frac{1}{2}S^x_j(t)\rangle -  2\langle S^x_i S^x_j(t) S^x_i\rangle.
\end{equation}
In \cite{Knap_PRL_2013} it was argued that such terms vanish for symmetry reasons since the Ising Hamiltonian is invariant with respect to $S^x\to - S^x$. However, our goal here is to average over the GGE rather than the thermal ensemble and symmetries of the Hamiltonian are not immediately useful for that purpose. We can apply the following argument instead: even though the Jordan-Wigner-transformed Hamiltonian Eq. (\ref{eq:transverse_Ising}) does not conserve the total number of Fermionic $c$-particles, it does conserve the particle parity. This can be seen from the fact that particle parity operator 
\begin{equation}
    P=\exp\left(\ii\pi\sum_{j=1}^L c^\dagger_j c_j\right)
\end{equation}
commutes with the Hamiltonian, $[H,P]=0$. Therefore, each eigenstate is a superpositions of either states with only even ($0,2,4,6,...$) or only odd ($1,3,5,7,...$) number of $c$-particles. By representing the products of $S^x$ operators by their Jordan-Wigner $c$-operator analogues,
\begin{equation}
    S^x_j=\frac{1}{2}\prod_{l<j}\exp\left(i\pi c^\dagger_l c_l\right) \left(c^\dagger_j + c_j\right)=\frac{1}{2}\prod_{l<j}\left( 1 - 2 c^\dagger_l c_l \right) \left(c^\dagger_j + c_j\right)
\end{equation}
it is obvious that they take even-parity states into odd and vice-versa, such that
\begin{equation}
    \langle \psi|S^x_j(t)|\psi\rangle = 0
\end{equation}
for any eigenstate $|\psi\rangle$ of $H$, and similarly for the three-operator average, hence the GGE averages of such operators vanish identically. Therefore, odd-$S^x$ terms vanish and many-body Ramsey protocol measurement Eq. (\ref{eq:M}) results in
\begin{equation}
 M_{ij}(0, \pi/2, t)=G_{ij,\mathrm{GGE}}^{xx}(t)\;.
\end{equation}
There is a subtlety involved concerning the symmetry-broken (anti)ferromagnetic phase: in the limit $L\to\infty$ some quasiparticle levels may become degenerate, leading, in particular, to two degenerate ground states, one with odd and one with even number of $c$-particles. Then symmetry-broken ferromagnetic state is possible with non-vanishing $\langle S^x\rangle\neq0$, but this does not happen for finite systems, which would be relevant in experiments.

\section{Qualitative behavior of $C^{xx}$ from semiclassical picture}

Spin-spin correlation functions $C^{xx}(r,t)\equiv C^{xx}_{i,i+r}(t)$ have been analyzed semiclassically, both in equilibrium at finite temperatures, and in non-equilibrium steady states following quantum quenches \cite{Sachdev_PRL_1997,Rieger_PRB_2011}. The semiclassical approach is based on treating the quasiparticles $\gamma_j^\dagger,\gamma_j$, introduced in Appendix A, as classical particles with populations $\langle N_j \rangle$, fixed by the conservation laws. Two limiting cases of $h=0$ and $|h|\gg |J|$ are explicitly considered, and the results are extended to the whole ordered ($|h|<|J|$) and paramagnetic ($|h|>|J|$) phases, respectively. Below we quote some results for the limit $L\to\infty$, when wavenumber $-\pi \leq k < \pi$ is a good quantum number and can be used to label the excitations with the corresponding dispersion relation 
\begin{equation}
    \epsilon_{k}=2\left(J^{2}+h^{2}-2Jh\cos\left(k\right)\right)^{1/2},
    \label{eq:quasipart_dispersion}
\end{equation}
and velocity $v_k=d \epsilon_k / dk$. Semiclassical treatment has been extended to finite systems \cite{Rieger_PRB_2011}, but for the case of large $L$ it suffices to use the $L\to\infty$ results.

In the $J>0$, $h=0$ ferromagnet, the quasiparticles diagonalizing Eq. (\ref{eq:H_TFI_1}) are the propagating domain walls with all the spins pointing in either positive or negative $x$ direction. The semiclassical picture thus consists of $\langle N_k \rangle$ excitations, which are domain walls traveling with velocities $v_k$ for each $-\pi < k < \pi$. In this simplified picture, for a given initial distribution of classical domain walls, the correlation function $C^{xx}(r,0)=\pm 1$, depending on how many quasiparticles reside on the interval $0<x<r$. Moreover, for $t>0$, the correlation function changes sign every time a quasiparticle enters or leaves this interval. Thus, to obtain $C^{xx}(r,t)$ one needs to average over every possible quasiparticle configuration. Following such probabilistic reasoning \cite{Sachdev_PRL_1997}, the ferromagnetic correlation function evaluates to
\begin{equation}
    C^{xx}(r,t)\sim\exp\left(-\int_{-\pi}^{\pi}\frac{dk}{\pi}\langle N_{k} \rangle \left|r-v_{k}t\right|\right).
\end{equation}

In the paramagnetic case $|h|>|J|$ the quasiparticles are single flipped spins in the $y$ direction with the same dispersion Eq. (\ref{eq:quasipart_dispersion}). Here the analysis is more involved since the correlation function $C^{xx}$ does not simply translate into the number of trajectories crossed in a space-time interval but also involves an oscillatory terms proportional to the modified Bessel function of the second kind,
\begin{equation}
    C^{xx}(r,t)\sim K_{0}\left(2\frac{|h|-J}{\sqrt{J|h|}}\left(r^2-J|h|t^2\right)^{1/2}\right)\exp\left(-\int^{\pi}_{-\pi}\frac{dk}{\pi}\langle n_{k} \rangle\left|r-v_{k}t\right|\right).
\end{equation}
\end{document}